\documentclass[aps,preprint,prb]{revtex4}

\usepackage{amsmath}
\usepackage[utf8]{inputenc}
\usepackage{url}

\begin{document}

\title{Condition for minimal Harmonic Oscillator Action}

\author{M.~Moriconi}
\email{mmoriconi@id.uff.br}
\affiliation{Instituto de Física, Universidade Federal Fluminense,
Campus da Praia Vermelha, Niterói, 24210-340, RJ, Brazil.}

\begin{abstract}
We provide an elementary proof that the action for the physical trajectory of the one-dimensional harmonic oscillator is guaranteed to be a minimum if 
and only if $\tau < \pi/\omega$, where $\tau$ is the elapsed time and $\omega$ is the oscillator's natural frequency.
\end{abstract}

\maketitle

\section{Introduction}

Hamilton's principle \cite{Fasano} establishes that the equations of motion for a mechanical system can be obtained as a necessary condition for the action to  
be locally stationary; that is, if the action is given by
\begin{equation}
S = \int_{t_1}^{t_2} L(q, \dot q, t) \,dt \, ,
\end{equation}
where $L$ is the Lagrangian of the system ($q$ representing the generalized coordinates), the condition $\delta S = 0$ implies the equations of motion. We are assuming that
the system is holonomic and that the constraints have been incorporated through a suitable choice of generalized coordinates.

The path that extremizes the action is often assumed to be the one that minimizes it, but this is not guaranteed by the first-order variation.  To determine whether the 
action is a maximum or a minimum one must analyze the second-order variation, similarly to what is done when studying critical points of real functions of one variable. 
In fact, it can be quite tricky to establish the minimality of the action for a given configuration. One of the cases where this analysis can be performed 
rigorously is for the one-dimensional harmonic oscillator (1D HO). A thorough analysis of this problem is provided by Gray and Taylor.\cite{Gray}

In order to understand the problem at hand, an analogy with a geometrical problem is very useful. Consider the paths joining two points, $P$ and $Q$, on 
the surface of a sphere of radius $R$. Without loss of generality, let us take the point $P$ to be on the north pole. If $Q$ is close to $P$, we know that the shortest 
path, as measured using the metric of the surface of the sphere, is given by the arc of the great circle joining these two points, and that this path is unique. As we move 
the point $Q$ farther away from $P$ this arc will keep growing, until $Q$ is on the antipode to $P$, the south pole. Once $Q$ is on the south pole, there is a surprising 
phenomenon---now there is an infinite number of arcs joining $P$ and $Q$ with minimal length.  For very good reasons, therefore, the point $Q$ is called the {\em kinetic focus}.  
We should note that the kinetic focus is more commonly known in the literature as the {\em conjugate point}, and we refer the reader to the book by Gelfand and Fomin \cite{Gelfand} 
for a readable treatment of this problem.

We see from this example that, given a point $P$ on the sphere, there exists a point $Q$, its kinetic focus, such that the geodesics that are shorter than the ones 
joining $P$ and $Q$ in fact minimize the path length.  (The term \textit{geodesic} here refers to curves that satisfy the geodesic equation and are not, necessarily, those 
with the shortest path length.)  For geodesics that are longer than those for the kinetic focus, it is no longer guaranteed that they will minimize the path length.  As a 
matter of fact, we know that an arc of a great circle with length greater than $\pi R$ is not the shortest path, even though it satisfies the geodesic equation.

In the action principle we are looking for paths that extremize the action, and we can ask a similar question: given a solution of the equations of motion---the true physical 
motion of a mechanical system---what are the conditions that will guarantee that this solution minimizes the action?  Notice that the kind of boundary conditions that will be 
imposed here are crucial.  We will study the case where the space-time coordinates of the initial and final points are given, which is necessary in writing the action principle. 

This question was first analyzed by Legendre,\cite{Goldstine} who thought he had solved it by looking at the second variation of the action $\delta^2 S$. We know that in order 
to guarantee that we have a minimum for the action, we must have $\delta^2 S > 0$. Legendre's conclusion for one-dimensional systems was that, for a Lagrangian of the 
form $L(q, \dot q, t)$, a necessary condition for the second variation of the action to be non-negative should be $\partial^2 L/ \partial \dot q^2\geq 0$. 

Legendre's condition is certainly necessary, for, intuitively, we can imagine varying the solution of the equations of motion by adding a small perturbation that changes 
the potential energy by a very small amount while changing the kinetic energy by a large amount. If $\partial^2 L/ \partial \dot q^2$ can be negative, this perturbation allows 
us to reduce the value of the action. Only later, with the work of Jacobi,\cite{Goldstine} it was understood that one needed to be careful about the existence of the kinetic focus.

For the harmonic oscillator there's a similar phenomenon to the problem of geodesics on a sphere. If we take the starting point to be $q_0 = 0$,  corresponding to the potential energy minimum, at $t_0 = 0$, and the end point to be $q_1$, arbitrary, which is reached 
at time $\tau < \pi/\omega$, then there's only one solution for the equations of motion. On the other hand, if we choose $\tau = \pi/\omega$, which corresponds to half 
of the period of oscillation, we see that there is an infinite number of solutions connecting $q_0 = 0$ at $t_0 = 0$ and $q_1 = 0$ at $t_1 = \pi/\omega$, namely, $q(t) = A \sin(\omega t)$ for any $A$. As shown by Gray 
and Taylor \cite{Gray}, it is possible to vary the path in such a way that the second variation of the action vanishes. and and in the more general case where the starting 
and end points are not at the minimum of the potential energy, as discussed by Gray and Taylor \cite{Gray}, to decide if we have a minimum or not. 

We will show now that the solutions of the equation of motion of the 1D HO connecting space-time points separated by a time interval less than $\pi/\omega$, provides, 
in fact, a true minimum of the action.  The Lagrangian for the 1D HO is given by
\begin{equation}
L = \frac{1}{2}m{\dot q}^2 - \frac{1}{2}kq^2 \, .
\end{equation}
Suppose we are given conditions $q(t_i)$, $i = 1, 2$, and that we find a unique solution for the equations of motion compatible with them, which we call $q_{\star}(t)$. We 
should note that for such boundary conditions it is not always the case that the solution is unique, as discussed by Gray. \cite{ScholarpediaGray} Without loss of generality 
we take $t_1 = 0$ and $t_2 = \tau$. The question is, does this solution minimize the action for the 1D HO?

Normally, one approaches this question by writing an arbitrary function $q(t)$ as $q(t) = q_{\star}(t) + \eta(t)$, where $\eta(t_i) = 0$, $i = 1,2$, and then computing $S[q(t)]$ 
as
\begin{eqnarray}
S[q(t)] &=& \frac{1}{2}\int_{t_1}^{t_2} \left[m(\dot q_\star + \dot \eta)^2 - k(q_\star + \eta)^2 \right] dt\nonumber \\
&=& \frac{1}{2}\int_{t_1}^{t_2} \left( m{\dot q_\star}^2 - kq_\star^2 \right) dt + \frac{1}{2}\int_{t_1}^{t_2} \left( m{\dot \eta}^2 - k\eta^2 \right) dt \nonumber \\
&=& S[q_\star(t)] + S[\eta(t)] \, ,
\end{eqnarray}
where we have performed an integration by parts (the total derivative term going to zero) and made use of the equation of motion.  In order to show that $S[q_\star(t)]$ is indeed 
the minimum action we must show that $S[\eta]\ge 0$ for non-zero $\eta(t)$, and equal to zero if and only if $\eta(t) = 0$. The traditional way to show this\cite{Gray} is to expand
$\eta(t)$ in Fourier series and then require $\delta^2 S>0$.  Due to the fact that $\eta(t_1) = \eta(t_2) = 0$, we can expand $\eta(t)$ in terms of $ \sin(n \pi t/\tau)$. The condition 
one obtains at the end is that $S[\eta(t)] \geq 0$ for any $\eta(t)$, if and only if
\begin{equation}
\tau < \frac{\pi}{\omega} \, , \label{key}
\end{equation}
where $\omega^2 = k/m$ is the oscillator's natural frequency.

This approach offers a very powerful method, but it requires the use of heavy mathematical machinery in order to establish the simple, but fairly important, condition 
for $\tau$.  Gray and Taylor \cite{Gray} present a derivation for the 1D and 2D HO using the Fourier series method and a derivation using the general theory of kinetic focus.  
In this note we establish Eq.~(\ref{key}) using only elementary calculus.

Following a beautiful trick by Lax in a related variational problem,\cite{Lax}  let us write $\eta(t) = u(t) \sin(\pi t/\tau)$, with $u(t)$ arbitrary, since the sine 
function guarantees that $\eta(t)$ vanishes for $t = 0$ or $\tau$.  Using this function for $\eta(t)$ and defining $\phi = \pi t/\tau$, we find
\begin{eqnarray} 
S[\eta(t)] &=& \frac{1}{2}\int_0^\tau \left[ m \left( \dot{u} \sin\phi + u \frac{\pi}{\tau} \cos\phi \right)^2 - ku^2 \sin^2\phi \right] dt \nonumber \\
&=& \frac{1}{2}\int_0^\tau \left[ m \dot{u}^2 \sin^2\phi + m u^2 \left(\frac{\pi}{\tau}\right)^2 \cos^2\phi \right. \nonumber \\
&\ &  \qquad\qquad  \left. +2\frac{m \pi}{\tau} u \dot{u} \sin\phi  \cos \phi  - ku^2 \sin^2\phi \right]dt \, .
\end{eqnarray}
We can then write
\begin{equation}
2\frac{m\pi}{\tau}u \dot u \sin\phi  \cos\phi = \frac{m\pi}{\tau}\left[\frac{d}{dt}\left( u^2 \sin\phi  \cos\phi\right) - \frac{\pi}{\tau} u^2 \cos^2\phi + 
\frac{\pi}{\tau} u^2 \sin^2\phi \right] \, ,
\end{equation}
allowing us to integrate by parts, so that the action becomes
\begin{equation} 
S[\eta] = \frac{1}{2} \int_0^\tau \left[ m\dot{u}^2 \sin^2\left( \pi \frac{t}{\tau}\right) + m\left(\frac{\pi^2}{\tau^2} - \omega^2 \right)
u^2 \sin^2\left(\pi \frac{t}{\tau}\right) \right] dt \, ,
\label{action_eta}
\end{equation}
where we have discarded the total derivative term, since it vanishes at $t = 0$ and  $t = \tau$.

The expression for the action given in Eq.~(\ref{action_eta}) allows us to find the necessary and sufficient conditions to guarantee that $S[\eta]$ is positive 
for all non-zero $\eta$. If we choose $u(t) = \kappa$, a constant, we obtain
\begin{equation}
S[\eta] = \frac{m}{2}\left(\frac{\pi^2}{\tau^2} - \omega^2 \right) \kappa^2  \int_0^\tau \sin^2\left(\pi \frac{t}{\tau}\right) dt =  \frac{m}{2}\left(\frac{\pi^2}{\tau^2} - 
\omega^2 \right) \kappa^2 \frac{\tau}{2} \, ,
\end{equation}
which, in order to be positive, requires
\begin{equation}
\frac{\pi^2}{\tau^2} - \omega^2 > 0\quad \implies \quad \tau < \frac{\pi}{\omega} \, ,
\end{equation}
establishing the fact that the Eq.~(\ref{key}) is a necessary condition. To establish its sufficiency, we note that if the Eq.~(\ref{key}) is satisfied, then the integrand 
in Eq.~(\ref{action_eta}) is positive definite.  Since the action is the sum of positive definite terms, its minimum value is obtained when each term is $0$; that is, 
for $u(t) =0$ and $S[q(t)] = S[q_\star(t)]$.  The final result is that the action is a minimum for time intervals less than half of the harmonic oscillator period; for greater 
time intervals the action integral is a saddle point, as defined by Gray and Taylor. \cite{Gray}
 
\begin{acknowledgments}
The author would like to thank the anonymous referees and the editor, whose comments helped improve this paper considerably, and to Nivaldo Lemos, for a critical reading.
\end{acknowledgments}

\end{document}